\documentclass[twocolumn]{aastex631}

\usepackage{hyperref}
\usepackage{booktabs}
\shorttitle{Evaluation of the effectiveness of sonification for time series data exploration}
\shortauthors{Guiotto Nai Fovino et al.}
\graphicspath{{./}{figures/}}

\begin{document}

\title{Evaluation of the effectiveness of sonification for time series data exploration}

\correspondingauthor{L. Guiotto Nai Fovino}
\email{lucrezia.guiottonaifovino@phd.unipd.it}

\author{L. Guiotto Nai Fovino}
\affiliation{University of Padua, Via Venezia 8, Padua (Italy)}

\author{A. Zanella}
\affiliation{Istituto Nazionale di Astrofisica, Vicolo dell'Osservatorio 5, 35137 Padova (Italy)}

\author{M. Grassi}
\affiliation{University of Padua, Via Venezia 8, Padua (Italy)}

\begin{abstract}
Astronomy is a discipline primarily reliant on visual data. However, alternative data representation techniques are being explored, in particular ``sonification'', namely, the representation of data into sound. While there is increasing interest in the astronomical community in using sonification in research and educational contexts, its full potential is still to be explored.
This study measured the performance of astronomers and non-astronomers to detect a transit-like feature in time series data (i.e., light curves), that were represented visually or auditorily, adopting different data-to-sound mappings. We also assessed the bias that participants exhibited in the different conditions.
We simulated the data of 160 light curves with different signal-to-noise ratios (SNR). We represented them as visual plots or auditory streams with different sound parameters to represent brightness: pitch, duration, or the redundant duration \& pitch. We asked the participants to identify the presence of transit-like features in these four conditions in a session that included an equal number of stimuli with and without transit-like features.
With auditory stimuli, participants detected transits with performances above the chance level. However, visual stimuli led to overall better performances compared to auditory stimuli and astronomers outperformed non-astronomers. Visualisations led to a conservative response bias (reluctance to answer  “yes, there is a transit”), whereas sonifications led to more liberal responses (proneness to respond “yes, there is a transit”).
Overall, this study contributes to understanding how different representations (visual or auditory) and sound mappings (pitch, duration, duration \& pitch) of time series data affect detection accuracy and biases.
\end{abstract}


\section{Introduction}
\label{sec:introduction}

Data sonification is described as “the transformation of data relations into perceived relations in an acoustic signal for the purpose of facilitating communication or interpretation” \citep{Kramer1999}. Simple examples of sonification involve binary messages, such as the bell sound we hear in modern cars when driving without seatbelts. However, more sophisticated sonifications, that transform more complex data such as numerical quantities into sound, have been under development for decades. Data sonification is a valuable tool in scientific research, facilitating a deeper understanding of the intrinsic narratives present in sets of data, a benefit for both the scientific community and the general public \citep{Sawe2020}. Sound can offer benefits in data analysis thanks to its multi-dimensional nature - pitch, loudness, duration, and timbre, among others - harnessing the sophisticated capabilities of the human auditory system. Our ability to process multiple sound streams and recognise patterns \citep{Bregman1990, Hermann2011, Sawe2020}, decode melodic contours, and notice slight variations in frequency, enhances the potential to uncover significant patterns embedded in the data \citep{Kramer2010}. These qualities make sonification particularly interesting for conveying time-dependent information, such as those collected through monitoring (e.g., see the sonification of electroencephalograms of epileptic seizures, \citealt{Matinfar2023, Supper2012}.
Traditionally, astronomy was studied and disseminated through visual means, such as images, animations, and graphs. Yet, most of the Universe's content (dark matter and dark energy) does not emit or absorb light. Even the matter that does often emits light at frequencies outside the human visible range or is too faint to be seen directly. Finally, data collected at the telescope are not collected as images, but rather as digital numbers that researchers turn into visuals (e.g., \citealt{Varano2023, GuiottoNaiFovino2023}). Existing reviews outlined the usefulness of astronomical sonification in both scientific and educational contexts \citep{Hermann2011, Zanella2022}. The astronomical community is increasingly turning to sound as a medium for data representation, as evidenced by the numerous contributions to the Data Sonification Archive \citep{Lenzi2020, Lenzi2021, Harrison2021, Zanella2022}. Sonification efforts aimed to enhance accessibility in public engagement, education and research activities, especially for those with visual impairments \citep{Bardelli2022, GarciaBenito2022, Harrison2022, Hyman2019}. It also proved to be a promising tool for exploring complex \citep{Cooke2019, Sawe2020} and multidimensional data sets and for monitoring time-series data without the need for visual attention \citep{Guttman2005}.
One promising application of sonification in time series data is the analysis of light curves. For example, \cite{DiazMerced2013} showed that sonification, when added to the visual display of noisy unidimensional data, increased sensitivity to events that might otherwise go unnoticed by the human eye. A common technique to sonify light curves is the so-called parameter-mapping sonification. In parameter-mapping sonification, it is necessary to establish a correspondence between the dimensions of the data and auditory features. Although it provides greater flexibility compared to other simpler types of sonification (e.g. the bell sound that we hear in cars when someone is not wearing a seatbelt is always the same, and we cannot identify who is not wearing the seatbelt), the design of individual mappings needs careful consideration. The most common sound dimension mapped to data is pitch \citep{Dubus2013}. This applies to the current literature on parameter-mapping sonification in astronomy as well \citep{Cooke2019}.
Several studies on the sonification of light curves and other time series data have been performed in recent years \citep{Cooke2019, DiazMerced2013, DiazMerced2008}. The most recent experimental study used the default Astronify\footnote{Astronify \citep{Brasseur}: \href{https://astronify.readthedocs.io/en/stable/}{Read The Docs link}} sonification algorithm, which maps brightness to pitch within a one-octave frequency range (440-880Hz, \citealt{TuckerBrown2022}). In the experiment of \cite{TuckerBrown2022}, the star's brightness was mapped as a function of time. The success rate in identifying signals in visual plots, sonification, and a combination of both was analyzed. Participants had to identify the presence of transits of different signal-to-noise ratios (SNRs). The results showed a consistently better performance for visual and plots combined with sonification in comparison to sonification only, especially for medium and low SNRs. Astronomers performed almost at ceiling for visual plots and were generally better than non-astronomers. Interestingly, sonification enabled the detection of signals with high SNR also for users with little to no experience in sonification \citep{TuckerBrown2022}. 
Building upon the investigation of \cite{TuckerBrown2022},  we investigated the ability of astronomers and non-astronomers to detect light curves of various SNRs. We tested three different sound mappings (pitch, duration, duration \& pitch) and compared them to the classic visual representation of data. Previous studies \citep{Cooke2019, DiazMerced2008, TuckerBrown2022} have represented time series data with sound by mapping brightness to pitch so that a drop in brightness (i.e., the dip of a  transient-like light curve) is associated with low-frequency sounds. We tested the reversed pitch association so that a decrease in brightness (i.e., the dip in a light curve) coincides with high-frequency sounds that are usually perceived as more salient than low-pitch ones \citep{Haas1996}. 
We also tested whether sound duration could be used as an alternative to pitch. Human sensitivity to sound’s duration is high, and hearing is, among the senses, the one with the highest sensitivity for duration \citep{Keetels2012}. Finally, we tested a redundant sonification with the simultaneous manipulation of sound frequency and sound duration. Unlike previous works \citep{TuckerBrown2022}, we did not investigate the performance only, but also the way sonification and visualization affected the response bias of the participants. The response bias reveals whether the participants produce more "yes, there is a transit" or "no, there is no transit" responses regardless of the presence (or absence) of the signal. It is the spontaneous tendency of the participant to use one (or the other) response. This is routinely used in the psychoacoustics domain when analyzing signal-in-noise detection experiments \citep{Green1966}.

The paper is structured as follows: in Section \ref{sec:method} we present the method that we adopted to recruit the participant, build the experiment and sonify the data; in Section \ref{sec:results} we report the statistical results of the experiment; in Section \ref{sec:discussion}, we discuss the results and their implications, finally in Section \ref{sec:conclusions} we summarize and conclude.

\section{Method}
\label{sec:method}

\subsection{Participants and apparatus}
\label{subsec:participants}
We recruited 58 participants for the experiment. 23 were astronomers with an average age of 42 years old, ranging from 28 to 64 years old. 35 participants were non-astronomers, with average age 36 years old, with an age range of 18 to 64 years old. All participants reported good hearing and normal or corrected-to-normal vision through the use of glasses or contact lenses. All participants were guaranteed anonymity.

We created the experiment using plugins from the jsPsych library \citep{DeLeeuw2015} and we administered it online through JATOS (v. 3.6.1). We sent the link to the experiment to the recruited participants, who accessed it with their preferred device. The majority of the participants used a laptop, except seven of them who used a tablet and one who used a mobile phone. The participants were free to adjust the intensity of the sonifications to their preference, and they were encouraged to use headphones.

\subsection{Synthetic light curves and adopted mappings}
\label{subsec:light_curves}
The experiment presented the data of 160 synthetic light curves translated into sonifications and visual plots, which will be called auditory and visual stimuli hereafter. We produced univariate, synthetic light curves with a customised IDL code, in the form of flux as a function of time. Each simulated light curve consisted of a time series of 80 evenly sampled data points. We added transit-like features as gradual drops in brightness, adopting Gaussian functions with specific full width at half maximum ($\mathrm{FWHM = 10}$, arbitrary units) and depth ($\mathrm{d = 20}$, arbitrary units). The simulated light curve could contain one or no transit. We added random noise extracted from a Gaussian distribution to achieve SNR of 5, 10, 20, or 40. 

\begin{figure}
    \begin{interactive}{animation}{figures/SignalSNR40.mp4} 
    \includegraphics[width=0.46\textwidth]{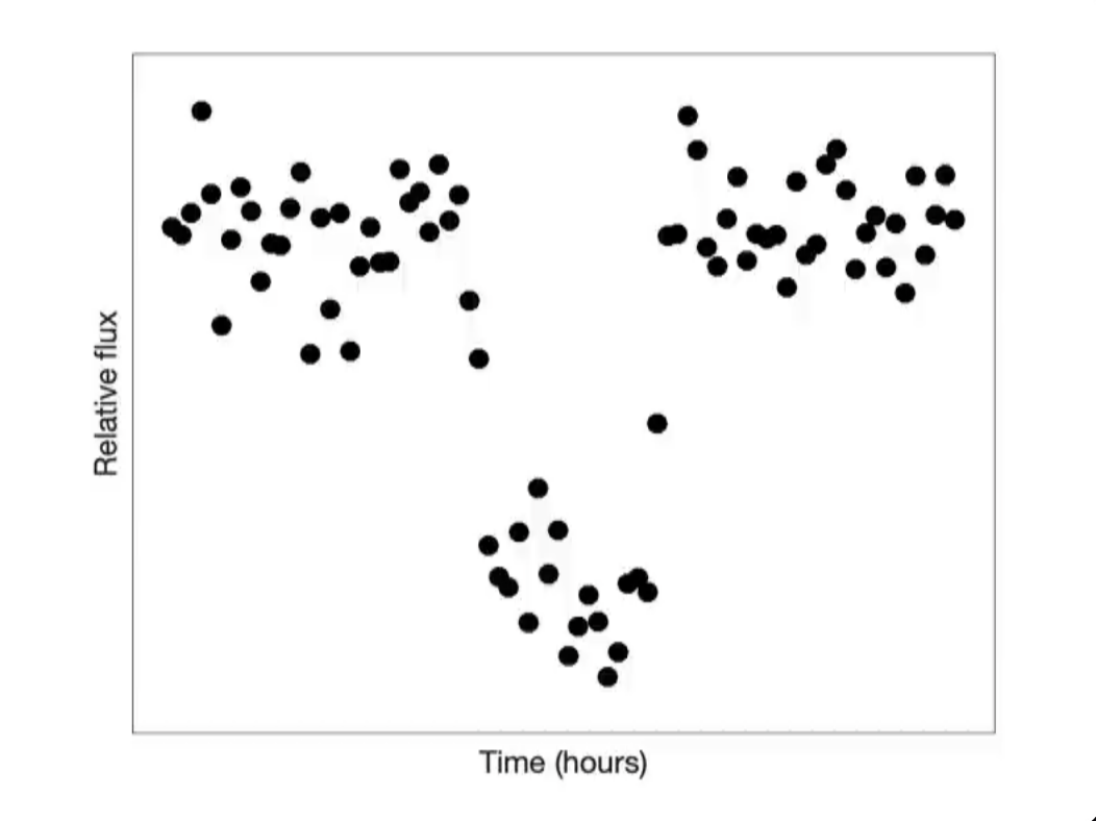}
    \end{interactive}
    \caption{Examples of the visual and sound representations of synthetic light curves adopted in the experiment. A light curve with SNR = 40 is shown. 
    \textbf{The three sonifications of this plot, where the relative flux is mapped to pitch, duration, or duration and pitch, are available both as an animation of this figure (in the HTML version of this article) and at this \href{https://osf.io/qy9ch/?view_only=10849f20b73d4d3b8a9127b81c11ea44}{link}. Relative flux decreases are mapped to increasing pitch and longer sound duration.} We did not plot axis tick marks on the plots to have visual stimuli consistent with our sonifications that did not contain analogous auditory marks. This was also useful to reduce the possible confusion of participants not used to interpret visual stimuli such as graphic plots. The reported units (e.g., hours) are simply indicative. 
    All stimuli used and other data are available in the \href{https://osf.io/k7w5h/?view_only=691aebb6f15b431299660d5d41487e01}{OSF repository link} 
    \dataset[(DOI 10.17605/OSF.IO/K7W5H)]{https://doi.org/10.17605/OSF.IO/K7W5H}.}
    \label{fig:light_curves}   
\end{figure}

\begin{figure}
    \begin{interactive}{animation}{figures/NoSignalSNR40.mp4} 
    \includegraphics[width=0.46\textwidth]{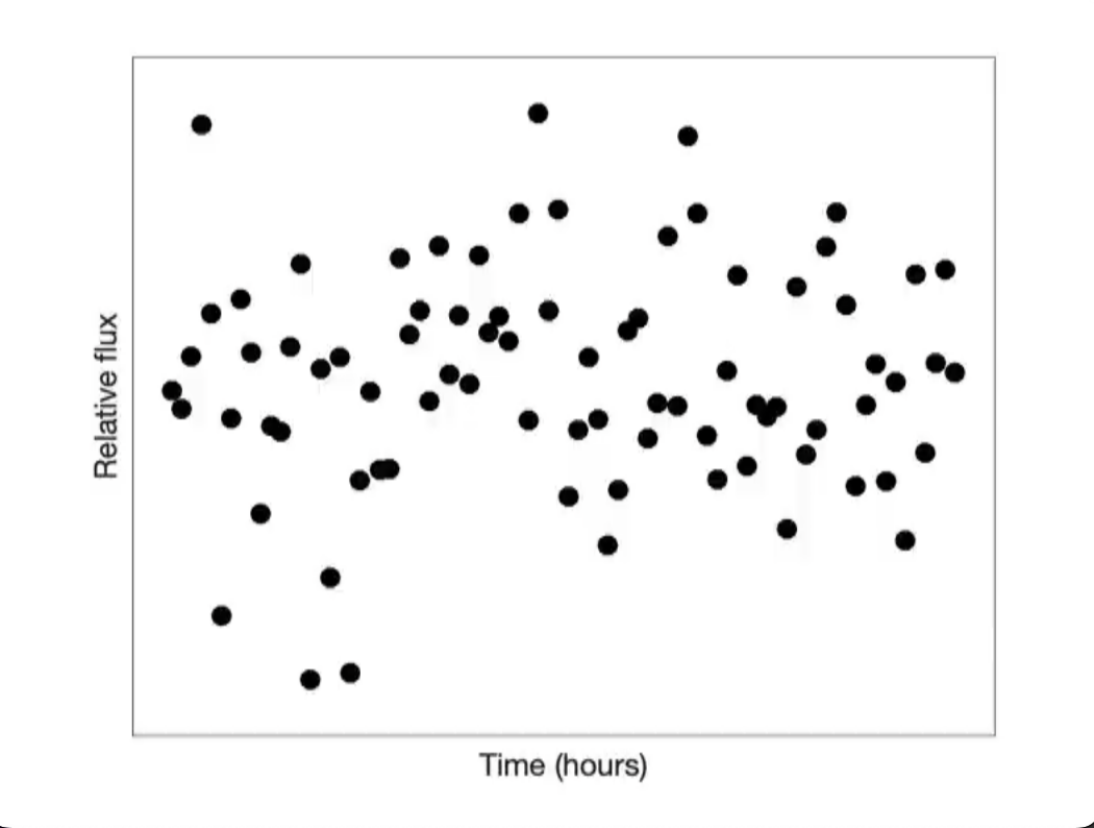} 
    \end{interactive}
    \caption{Examples of the visual and sound representations of synthetic light curves adopted in the experiment. A light curve without transit (i.e., no signal). 
    \textbf{The three sonifications of this plot, where the relative flux is mapped to pitch, duration, or duration and pitch, are available both as an animation of this figure (in the HTML version of this article) and at this \href{https://osf.io/v9ht4/?view_only=e457bbf25444468bb79ae1f779dda016}{link}. Relative flux decreases are mapped to increasing pitch and longer sound duration.} As in Figure \ref{fig:light_curves}, we did not plot axis tick marks on the plots to have visual stimuli consistent with our sonifications that did not contain analogous auditory marks.
    All stimuli used and other data are available in the \href{https://osf.io/k7w5h/?view_only=691aebb6f15b431299660d5d41487e01}{OSF repository link} 
    \dataset[(DOI 10.17605/OSF.IO/K7W5H)]{https://doi.org/10.17605/OSF.IO/K7W5H}.}
    \label{fig:light_curves_nosig}
\end{figure}

Each simulated light curve was presented to the participants in the form of visual or auditory stimulus. The visual stimuli were plots showing brightness as a function of time (Figures \ref{fig:light_curves} and \ref{fig:light_curves_nosig}). We did not plot axis tick marks on the plots to have visual stimuli consistent with our sonifications that did not contain analogous auditory marks. This was also useful to reduce the possible confusion of participants that are not used to interpret graphic plots.

The auditory stimuli represented the light curves as a stream of eighty tones. Three different types of data-to-sound mappings were used in the experiment to create the streams. In the first, flux was mapped to sound duration; in the second, flux was mapped to pitch; in the third, both mappings were applied together. For all mappings, the timbre was a sine wave and each tone of the streams was modulated in amplitude with an exponential ramp that attenuated the amplitude of the tone over its duration (i.e., the tones were not steady in amplitude). This attenuation made the tone’s perception more similar to tones we normally listen to in everyday life such as piano tones, guitar tones, and all tones that are characterised by a fast attack followed by a gradual decay. The onset and the offset of the tones were gated on and off with two 5-ms raised cosine ramps. Tones were synthesised at 44.1 kHz and 16-bit resolution and saved as mp4 files. Auditory streams had a duration of 20 seconds (pitch sonification streams) or about 20 seconds (duration and duration \& pitch sonification streams). 
When the flux was mapped to duration, the tone frequency was fixed at 565.6 Hz, whereas the tone duration ranged from 33 to 500 ms in logarithmic steps. Brighter data points had shorter durations, whereas dimmer data points (i.e., the dip of the light curve) had longer durations. When flux was mapped to pitch, the duration of each tone was fixed to 125 ms. The tone frequency ranged from 100 Hz to 3200 Hz in logarithmic steps. Brighter data points were associated with low-frequency tones, whereas darker data points (such as the dip of the light curve) were associated with higher-frequency tones. Finally, for the duration \& pitch mapping, both the aforementioned mappings were used together, hence the tones of the stream varied both in pitch and duration.

\subsection{Procedure}
\label{subsec:procedure}
The experiment was structured as follows. First, participants were required to provide their informed consent. Then, we collected basic information necessary for the study, including gender, age, employment, device used for the experiment, and whether or not they were using headphones. Participants were asked to self-evaluate their proficiency in understanding plots and their musical ability on a 4-point Likert scale (none, poor, good, excellent). They were given instructions on how to complete the task and were given an example of how a signal or its absence appeared in all the experimental conditions (duration, pitch, duration \& pitch, visual) with SNR of 40 (strong signal) and 10 (weak signal). One example of data without signal (i.e., no transit feature) was also presented both for the auditory and the visual mappings. These examples were presented to help participants understand and become acquainted with the task. Then the actual experiment began. 
The participants were presented with 160 light curve sonifications and visualisations in a randomised order. Eighty stimuli did not include a signal (no transit-like flux dip), whereas eighty included a signal (presence of the dip). In Table \ref{tab:stimuli} we summarise the stimuli that we presented to the participants. After the presentation of each stimulus, the participants pressed "Y" if they believed the signal was present, and "N" if they believed it was not. Participants did not receive feedback on the correctness of the response. 

\begin{table}
    \centering
    \begin{tabular}{c c c}
       \toprule
       Type of stimulus & Signal  & No Signal\\
        & With SNR = 5, 10, 20, 40 & \\
       \midrule
       Visual & 20  & 20 \\
       Pitch & 20 & 20 \\
       Duration & 20 & 20 \\
       Duration \& pitch & 20 & 20 \\
       \midrule
       Total & 80 & 80 \\
       \bottomrule
    \end{tabular}
    \caption{Summary of the stimuli presented to the participants.}
    \label{tab:stimuli}
\end{table}

\section{Results}
\label{sec:results}
\begin{figure*}
    \centering
    \includegraphics[width=0.9\textwidth]{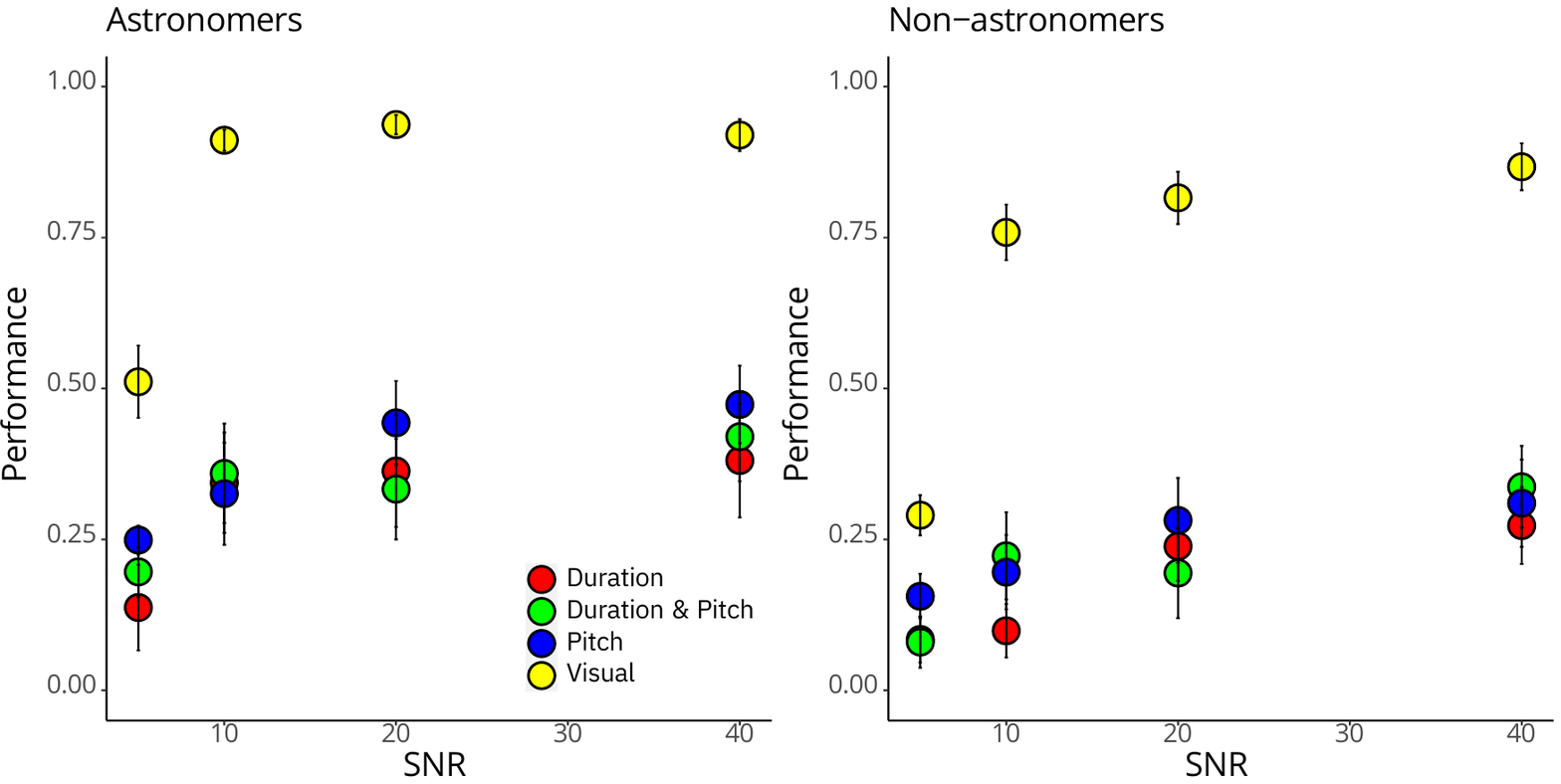}
    \vspace{-1cm}
    \caption{Performance as a function of SNR (\textbf{left panel}: astronomers, \textbf{right panel}: non-astronomers). Error bars represent the standard error of the mean. \textbf{Visual stimuli are shown in yellow, pitch stimuli in blue, duration stimuli in red, and duration \& pitch in green.}}
    \label{fig:performance}
\end{figure*}

\begin{figure*}
    \centering
    \includegraphics[width=0.9\textwidth]{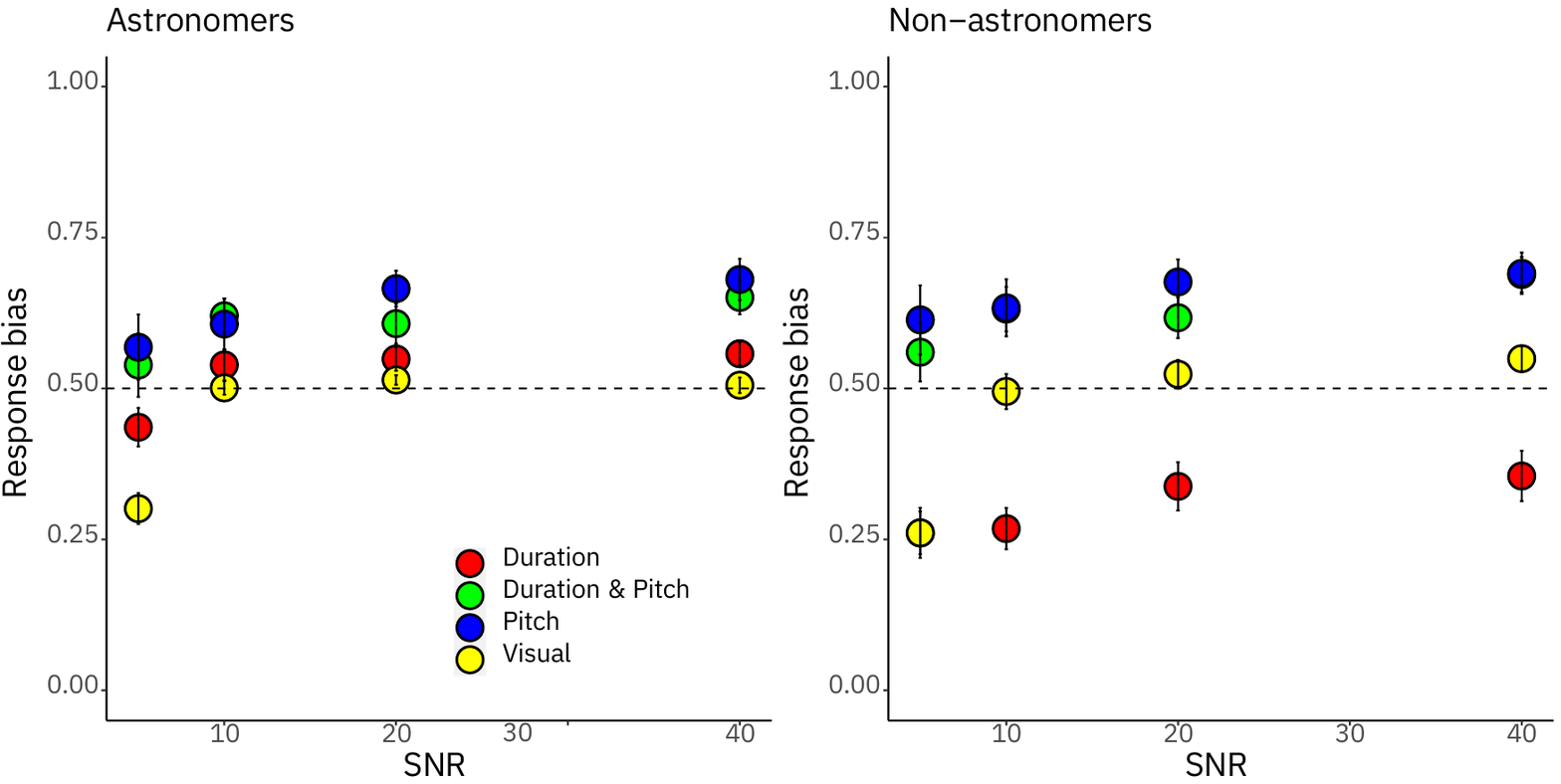}
    \vspace{-1cm}
    \caption{Response bias as a function of SNR. \textbf{Left panel}: results for astronomers. \textbf{Right panel}: results for non-astronomers. Error bars represent the standard errors of the mean. \textbf{Visual stimuli are shown in yellow, pitch stimuli in blue, duration stimuli in red, and duration \& pitch in green.} The dashed line shows no bias. If the data point is below 0.5, it means that participants were prone to respond ``No, the signal is absent'' for that stimulus. If the data point is above 0.5, it means that participants were prone to respond ``Yes, the signal is present'' for that stimulus.}
    \label{fig:bias}
\end{figure*}

We calculated the proportion of affirmative responses ("yes, the signal is present") for each participant and, separately, for each type of stimulus  (i.e., signal present/signal absent), modality of the stimulus (visual or auditory, which comprised pitch, duration, and duration \& pitch), and SNR ratio (Figure \ref{fig:performance}). These proportions were the “hits” (H, the participant responded “yes, there is a transit” and the light curve included a flux dip) and the “false alarms” (FA, the participant responded “yes, there is a transit” and the light curve did not include a dip). We calculated two indexes to represent two aspects of the participant’s response. These indexes were inspired by the works on the signal detection theory \citep{Green1966}. The first way to represent the participant’s responses is referred to as “performance”, which is the difference, for each given stimulus, between the hit and the false alarm rate of the participant: $\mathrm{P = H - FA}$. Performance is the net accuracy of the participant over his/her false alarm rate. Theoretically, performance can range from $\mathrm{P = -1}$ and $\mathrm{P = 1}$. When $\mathrm{P = -1}$ the participant made a perfect, but reversed, performance, such as a participant that reversed the meaning of the instructions received before the experiment. When $\mathrm{P = 1}$ the participant produced hits only and no false alarms. Finally, $\mathrm{P = 0}$ indicates that the participant cannot distinguish the stimuli including a signal (light curve dip) from those that do not include it. 
An alternative way to represent the participant’s responses is referred to as “bias” (B, Figure \ref{fig:bias}). This is the mean of the hits and the false alarms rate of the participant. The bias lets emerge the tendency of the participant to use more the “yes, there is a transit” response or the “no, there is no transit” response regardless of the performance. Bias ranges from 0 to 1. For B = 0 the participants responded always “no, there is no transit”, while for $\mathrm{B = 1}$ the participants responded always “yes, there is a transit”. $\mathrm{B = 0.5}$ means that the participant used both responses with a similar frequency.
In our experiment, an ideal respondent should have a high performance but a bias of 0.5, revealing that s/he distributed evenly the “yes” and the “no” responses (the experiment included 50\% signal trials and 50\% noise trials). In other words, they would produce many hits but no false alarms. Usually, stimuli that are easy to detect are associated with no response bias (B = 0.5). Bias becomes an interesting way to represent the responses in the case of stimuli that are difficult to detect (e.g., low SNR). For these stimuli it is possible to understand whether the participant is more prone to produce a “yes” or a “no” response, in other words, whether the participant had a liberal or a conservatory response criterion.
We calculated an analysis of variance (ANOVA) on performance considering the following within factors: type of stimulus (duration, duration \& pitch, pitch, visual) and SNR. The group (astronomers/non-astronomers) was also included as a between factor. This analysis investigates whether performance changed as a function of SNR, mapping used to represent the light curve data, and the group of participants, or subsets of these variables when combined in pairs or triplets. We show the results in Figure \ref{fig:performance}.
The ability of the participants to detect a transit-like feature increased as a function of SNR: F(3, 168)=51.19, $\mathrm{p < 0.0001}$. In general, the performance with the visual stimuli was better than with the auditory stimuli for any level of SNR: $\mathrm{F(3, 168)=95.49}$, $\mathrm{p < 0.0001}$. An interaction between stimulus type and SNR was also observed: $\mathrm{F(9, 504)=14.21}$, $\mathrm{p < 0.0001}$. This result revealed that performance increased differently for the various types of stimuli as a function of SNR. In the case of visual stimuli, the growth of performance was abrupt whereas in the case of sound stimuli, the growth of performance was more gradual. Astronomers had a performance averagely higher than non-astronomers: $\mathrm{F(1, 56)=4.78}$, $\mathrm{p = 0.03}$. No other statistical significance was observed. These findings are summarised in Table \ref{tab:anova_performance}.

We repeated the same calculations on the response bias. No overall difference was observed between astronomers and non-astronomers: $\mathrm{F(1, 56)=2.43}$, $\mathrm{p > 0.05}$. Bias changed as a function of the SNR: $\mathrm{F(3, 168)=51.19}$, $\mathrm{p<0.0001}$. Bias also changed for stimulus type: $\mathrm{F(3, 168)=30.19}$, $\mathrm{p <0.0001}$. Interestingly, bias was different for astronomers and non-astronomers as a function of the type of stimuli: $\mathrm{F(3, 168)=6.36}$, $\mathrm{p<0.00004}$. A visual inspection of the responses when represented as bias revealed that, for astronomers, the auditory conditions and the visual conditions elicited different biases (Figure \ref{fig:bias}). With visual stimuli astronomers showed no bias (and ceiling performance) for high SNRs. In contrast, for the lowest SNR they showed a positive performance but a tendency to be reluctant to produce a “yes” response: the performance was the result of several correct rejections and relatively few hits. In contrast, for almost all auditory stimuli, regardless of the performance, astronomers were more prone to use the “yes” response, also for the lowest SNR stimuli. For non-astronomers, the auditory stimuli using the duration as a mapping tended to gather a “no” response, whereas the rest of the stimuli (except one) tended to gather a “yes” response. 
Interestingly, when presented with visual stimuli with low SNR, both astronomers and non-astronomers showed a conservative response use and rarely responded “yes”. Noticeably, auditory stimuli with low SNR are less affected by this behaviour, and for these stimuli participants use the “yes” response more often (except for the responses to duration stimuli for non-astronomers). However, this more abundant use of the “yes” response did not lead to an increment in performance, and these responses are likely distributed equally among hits and false alarms. 

\begin{table*}
    \centering
    \begin{tabular}{p{4.7cm} c c c p{6cm}}
    \toprule
        Factor & F-value & Degrees of freedom & p-value & Effect Description \\
    \midrule
    SNR & 51.19 & (3, 168) & $<0.0001$ & Detection of transit increases with SNR \\
    Stimulus Type & 95.49 & (3, 168) & $<0.0001$ & Performance with visual stimuli is better than with auditory stimuli \\
    SNR $\times$ Stimulus Type & 14.21 & (9, 504) & $0<.0001$ & Performance growth differs for type of stimuli as a function of SNR \\
    Astronomers vs Non-astronomers & 4.78 & (1, 56) & 0.03 & Astronomers perform better on average than  non-astronomers \\
    \bottomrule
    \end{tabular}
    \caption{Summary of the ANOVA results calculated on the performance index.}
    \label{tab:anova_performance}
\end{table*}

\begin{table*}
    \centering
    \begin{tabular}{p{4.7cm} c c c p{6cm}}
        \toprule
    Factor & F-value & Degrees of freedom & p-value & Effect Description \\
    \midrule
    SNR & 51.19 & (3, 168) & p $<0.0001$ & Bias changed as a function of SNR \\
    Stimulus Type & 30.19 & (3, 168) & p $<0.0001$ & Bias changed for stimulus type \\
    Astronomers vs Non-astronomers $\times$ Stimulus Type & 6.36 & (3, 168) & p $<0.0004$ & Bias was different between astronomers and non-astronomers as a function of the type of stimuli \\
    Astronomers vs Non-astronomers & 2.43 &(1, 56) & p $>0.05$ & There was no difference between the average bias of astronomers and non-astronomers \\
        \bottomrule
    \end{tabular}
    \caption{Summary of the ANOVA results calculated on the bias index.}
    \label{tab:anova_bias}
\end{table*}

\section{Discussion}
\label{sec:discussion}
In the current study, we investigated the effectiveness of sonification in conveying information contained in time series data and whether different parameter mappings yielded different performances in the recognition of events. We also compared sonification with visualisation and examined whether the visual or auditory nature of the stimuli affected the participants' response bias. 
Our analysis revealed that the participants performed better with visual stimuli than with sonified stimuli at all SNR levels, albeit the difference decreased at low SNR both for astronomers and non-astronomers. Detection increased with higher SNR for all groups and for all conditions. 

\subsection{\textbf{Performance Analysis for Astronomers and Non-Astronomers}}
Astronomers showed notably higher efficiency than non-astronomers, especially in visual conditions at medium and high SNR levels, achieving near-ceiling performance. With sonified stimuli, astronomers also outperformed non-astronomers. Interestingly, their performance with auditory stimuli varied significantly with SNR, ranging from P $\sim$ 0.25 to 0.50.

Non-astronomers exhibited a similar trend as astronomers, although their peak performance for visual stimuli occurred at higher SNR values (SNR $\geq$ 40). Our results align partially with \cite{TuckerBrown2022}, showing lower overall performance but a similar pattern. However, a direct comparison of the two experiments is not possible because of the different methods used in the studies such as the number of signals per trial or the calculation of performance.

\subsection{\textbf{Bias Analysis in Response Strategies}}
The analysis of the bias index showed some differences between astronomers and non-astronomers. The use of the response in the two groups (i.e., being more liberal and producing more “yes” responses or more conservative and producing more “no” responses) was affected by SNR and stimulus type. A visual inspection indicated that astronomers showed no bias for high SNR visual stimuli because these stimuli were easy to discriminate. However, for low SNR, they were reluctant to respond "yes, there is a transit" and adopted a conservative response criterion: the positive performance emerged thanks to many correct rejections and a very small number of false alarms, and not thanks to hits. In contrast, astronomers had a tendency to respond "yes" to auditory stimuli regardless of SNR. Non-astronomers showed a similar behaviour except for the duration stimuli in which the positive performance emerged thanks to correct rejections and not thanks to hits. Finally, in general, low SNR visual stimuli led to conservative responses in both groups, but this was not seen with low SNR auditory stimuli.

\subsection{\textbf{Evaluation of Sonification Parameters}}
We found that sonifying data with pitch or duration yielded comparable results. This is interesting, as it suggests that duration mapping could be used for the sonification of time-series data with the same effectiveness as pitch, at least for astronomers. Using a redundant sonification that manipulated pitch and duration at the same time did not offer any benefit, and results gathered with the redundant sonification were comparable to the pitch-only condition in terms of performance and bias, suggesting that -when used together- pitch tends to be predominant. Sonification gathered positive performances, in the sense that participants (both astronomers and non) were able to detect transits above chance level even at the lowest SNR. However, visually inspecting time-series data is still the best option for astronomers and non-astronomers. This may be due to visual stimuli being easier to detect or, in alternative, due to the fact that we are more trained since childhood to interpret this type of stimuli (e.g., plots). This is emphasised even more for astronomers, who are used to inspect visual plots in their daily work and this translates into the ceiling performances when inspecting visual stimuli of medium and high SNR. The question remains whether the performance in the sonification cases could be improved with more training.
Our results suggest that inspecting data through sonification could lead to a more liberal response criterion. Here, participants produced more “yes” responses for this type of stimuli (with the exception of non-astronomers and duration-stimuli). This more liberal response criterion could be exploited to raise the low rate of “yes” responses that is observed for low SNR visual stimuli. In practice, we may wonder whether, by presenting audiovisual stimuli (i.e., combining visualization to sonification) the observer could benefit from both types of representation: the better accuracy of the visual inspection of the stimulus combined with the higher feeling to detect something delivered by auditory stimuli. A behaviour as such would raise false alarms but also genuine discoveries that, at the moment, are hidden in the conservative approach that observers show when looking at low SNR stimuli. 

\section{Summary and conclusions}
\label{sec:conclusions}
Data sonification leverages the human auditory system's ability to finely process and recognize sound patterns. Sonification can be used to enhance existing datasets or to represent additional dimensions in complex multidimensional datasets, making it particularly interesting for astronomy \cite{Cooke2019, DiazMerced2013, DiazMerced2008, TuckerBrown2022}. Our experiment tested three sonification mappings and compared participants' performance and response bias in identifying transits with those obtained using visual plots. Fifty-eight astronomers and non-astronomers took part in our study. We generated synthetic light curves and visualised them as plots or sonified them as streams of sounds, adopting different data-to-sound mappings (duration, pitch, or both).
Our results revealed that:
\begin{itemize}
\item Both astronomers and non-astronomers were able to exploit sonification to detect transit-like features. However, performance with visual stimuli was higher compared to sonified stimuli.
\item Astronomers outperformed non-astronomers in all conditions of the experiment. 
\item Visualization and sonification yielded different response biases: visualization yielded no bias (for $\mathrm{SNR \geq 10}$) whereas it yielded a conservative bias for $\mathrm{SNR < 10}$. In contrast, sonification gathered a more liberal response bias. The only exception was the response bias for duration stimuli in non-astronomers that yielded a conservative bias at all SNR.
\end{itemize}
Future developments of this project will need more controlled experimental conditions (e.g., all users use headphones). We will also test the impact of training participants about sonification and include people who are already sound-trained (e.g., musicians) in studies.

\section*{Acknowledgements}
The authors thank Martina Barbieri for data collection {, and the
    reviewer and data editor of the journal for their constructive
    feedback that improved the clarity of the manuscript}.
The stimuli, analysis scripts, and datasets used in this study are publicly available 
in the \href{https://osf.io/k7w5h/?view_only=691aebb6f15b431299660d5d41487e01}{OSF repository}:
\dataset[(DOI 10.17605/OSF.IO/K7W5H)]{https://doi.org/10.17605/OSF.IO/K7W5H} \citep{10.17605/osf.io/k7w5h}

\bibliography{new.ms}{}
\bibliographystyle{aasjournal}

\end{document}